\documentclass[journal]{IEEEtran}

\usepackage{cite}
\usepackage[pdftex]{graphicx}
\usepackage{xcolor, tikz, subcaption}
\usepackage{amsmath, amsfonts, bm, scalerel}
\usepackage[ruled, vlined, commentsnumbered, linesnumbered]{algorithm2e}
\usepackage{algorithmic}
\usepackage{array, multirow}
\usepackage{url, hyperref}

\newcommand{\etal}{{\em et al.}}       
\newcommand{\eg}{{\em e.g.}}           
\newcommand{\ie}{{\em i.e.}}           

\DeclareMathOperator*{\concat}{\scalerel*{\|}{\bigodot}}

\begin{document}
\title{Multi-Scale Neural Network for\\EEG Representation Learning in BCI}

\author{Wonjun~Ko,~Eunjin~Jeon,~Seungwoo~Jeong,~and~Heung-Il~Suk,~\IEEEmembership{Member,~IEEE}
\thanks{W. Ko, E. Jeon, and H.-I. Suk are with the Department of Brain and Cognitive Engineering, Korea University, Seoul 02841, Korea. S. Jeong and H.-I. Suk are with the Department of Artificial Intelligence, Korea University, Seoul 02841, Korea. Correspondence: hisuk@korea.ac.kr (Heung-Il Suk)}%
}\markboth{UNDER REVIEW}
{Shell \MakeLowercase{\textit{et al.}}: Bare Demo of IEEEtran.cls for IEEE Journals}

\maketitle

\begin{abstract}
Recent advances in deep learning have had a methodological and practical impact on brain--computer interface (BCI) research. Among the various deep network architectures, convolutional neural networks (CNNs) have been well suited for \emph{spatio-spectral-temporal} electroencephalogram (EEG) signal representation learning. Most of the existing CNN-based methods described in the literature extract features at a sequential level of abstraction with repetitive nonlinear operations and involve densely connected layers for classification. However, studies in neurophysiology have revealed that EEG signals carry information in different ranges of frequency components. To better reflect these \emph{multi-frequency} properties in EEGs, we propose a novel deep \emph{multi-scale neural network} that discovers feature representations in multiple frequency/time ranges and extracts relationships among electrodes, \ie, spatial representations, for subject intention/condition identification. Furthermore, by completely representing EEG signals with spatio-spectral-temporal information, the proposed method can be utilized for diverse paradigms in both active and passive BCIs, contrary to existing methods that are primarily focused on single-paradigm BCIs. To demonstrate the validity of our proposed method, we conducted experiments on various paradigms of active/passive BCI datasets. Our experimental results demonstrated that the proposed method achieved performance improvements when judged against comparable state-of-the-art methods. Additionally, we analyzed the proposed method using different techniques, such as PSD curves and relevance score inspection to validate the multi-scale EEG signal information capturing ability, activation pattern maps for investigating the learned spatial filters, and t-SNE plotting for visualizing represented features. Finally, we also demonstrated our method's application to real-world problems.
\end{abstract}

\begin{IEEEkeywords}
Active/Passive Brain--Computer Interface; Electroencephalogram; Deep Learning; Convolutional Neural Network; Motor Imagery; Steady-State Visually Evoked Potentials; Mental Fatigue; Seizure
\end{IEEEkeywords}

\IEEEpeerreviewmaketitle

\section{Introduction}
\label{sec: introduction}
\IEEEPARstart{B}{rain}--computer interface (BCI) \cite{blankertz2008optimizing} is an emerging technology that enables a communication pathway between a user and an external device (\eg, a computer) through the acquisition and analysis of brain signals. Then these signals are translated into commands that are understood by a device, such as a computer. Owing to its practicality, electroencephalogram (EEG)-based non-invasive BCIs are widely used \cite{schirrmeister2017deep, blankertz2008optimizing, lawhern2018eegnet}. Earlier, Aric\`{o} \etal~\cite{arico2018passive} categorized user-centered BCIs into two types, active/reactive and passive BCIs. In this paper, our focus is not only on active BCIs but also on passive BCIs. Generally, two types of brain signals such as \emph{evoked} and \emph{spontaneous} EEG are primarily considered for active/reactive BCIs \cite{zhang2019survey}. Evoked BCIs exploit unintentional electrical potentials reacting to external or internal stimuli. Examples of evoked BCIs include steady-state visually evoked potentials (SSVEP) \cite{lee2019eeg, nakanishi2015comparison} and event-related potentials \cite{lee2019eeg}. Additionally, spontaneous BCIs use an internal cognitive process such as event related desynchronization and event related synchronization (ERD/ERS) in sensorimotor rhythms, \eg, motor imagery (MI) \cite{lee2019eeg, cho2017eeg} induced by imagining movements in addition to physical movement. Well-known examples of passive BCIs include the use of sleep/drowsy EEG signals for sleep stage classification or identifying mental fatigue to alert a driver of a dangerous situation and seizure EEG patterns for onset detection to provide the patient with a warning of a potential seizure.

Generally, machine learning-based BCIs consist of five main processing stages \cite{lawhern2018eegnet}: (i) an EEG signal acquisition phase based on each paradigm, (ii) signal preprocessing (\eg, channel selection and band-pass filtering), (iii) feature representation learning, (iv) classifier learning, and finally (v) a feedback stage. Basically, most of machine learning-based BCI methods follow these processes, however, these methods need specific modification to classify a user's intention/condition for each different paradigm \cite{lawhern2018eegnet}. In other words, machine learning-based methods need to have \emph{prior} knowledge of different EEG paradigms \cite{blankertz2008optimizing, lee2019eeg, nakanishi2015comparison, lawhern2018eegnet, shoeb2009application}. Therefore, conventional machine learning-based BCIs have discovered EEG representations through extremely specialized approaches, \eg, a common spatial pattern (CSP) \cite{blankertz2008optimizing} or its variants \cite{suk2012novel, ang2008filter} for MI signals and a canonical correlation analysis (CCA) \cite{nakanishi2015comparison} for SSVEP signals decoding.

While hand-crafted feature representation learning has a pivotal role in a conventional machine learning framework \cite{blankertz2008optimizing, nakanishi2015comparison, shoeb2010application}, deep learning-based representation has had remarkable results in the BCI community \cite{schirrmeister2017deep, lawhern2018eegnet, zhang2019survey}. These deep learning-based methods have integrated a feature extraction step with a classifier learning step such that those steps are jointly optimized, thereby improving performance. Among various deep learning methods, convolutional neural networks (CNNs) have the advantage \cite{ko2018deep, lawhern2018eegnet, chollet2017xception} of maintaining the structural and configurational information in the original data. In this respect, developing a novel CNN architecture for EEG signal representation has taken a center stage in the BCI studies \cite{schirrmeister2017deep, supratak2017deepsleepnet, sakhavi2018learning, ko2018deep, waytowich2018compact, kwak2017convolutional, asif2019seizurenet, emami2019seizure, gao2019eeg}.

\begin{figure}[t]
	\centering
	\includegraphics[width=.96\linewidth]{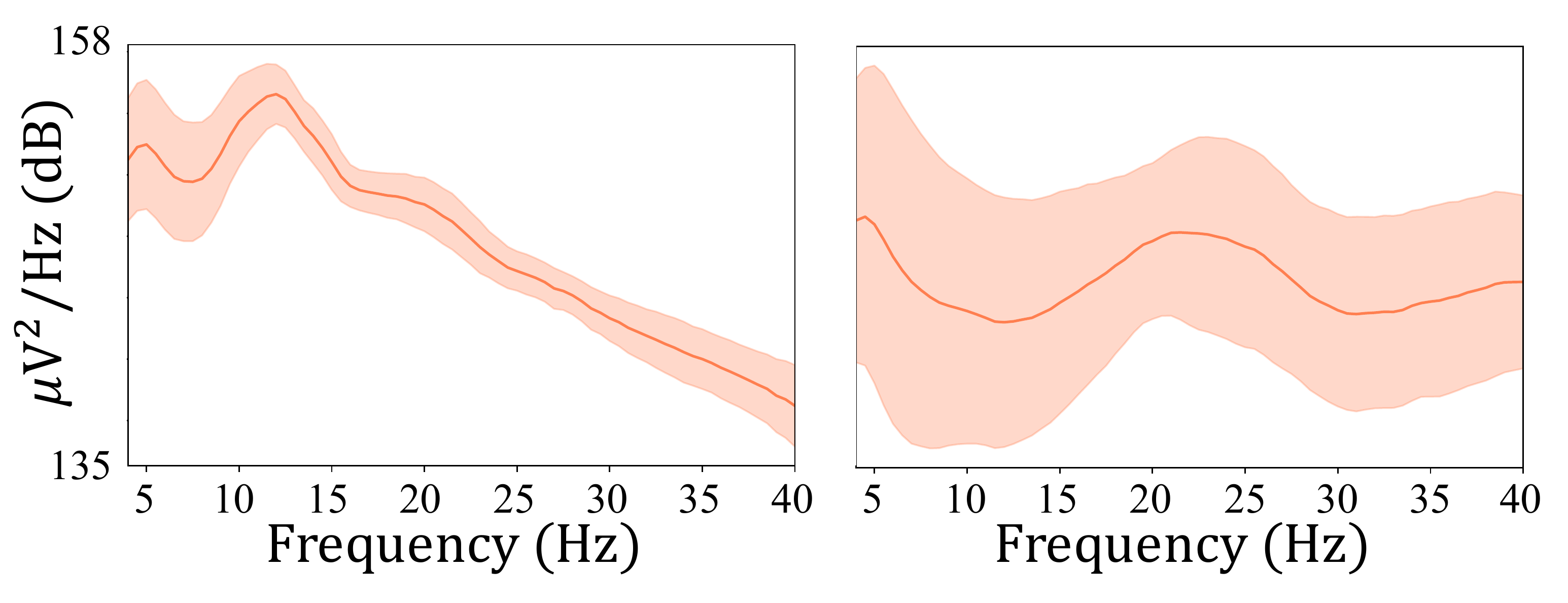}
	\caption{Power spectral density (PSD) curves of two different subjects' MI EEG samples. The solid red line denotes the mean PSD and the shaded region exhibits the standard deviation for all trials. Clearly, these two different subjects show quite different PSD patterns for the same paradigm (motor imagery).}
	\label{fig: psd_curves}
\end{figure}

However, some challenges still remain. First, existing CNN-based methods \cite{schirrmeister2017deep, ko2018deep, sakhavi2018learning, supratak2017deepsleepnet, emami2019seizure, gao2019eeg} are mostly comprised of stacked convolutional layers. In other words, those existing methods extract features sequentially. But, ignoring multiple ranges of spectral-temporal features can cause a critical problem because EEG signal features for different subjects \cite{jayaram2016transfer}, paradigms \cite{lawhern2018eegnet}, and types \cite{arico2018passive} are found in diverse ranges. For example, Fig. \ref{fig: psd_curves} depicts two different subjects' MI EEG power spectral density (PSD) curves. Clearly, these two plots have different distributions from each other even though these PSDs are estimated by the same task. Therefore, it is important to capture multi-scale spectral information in EEGs for \emph{general use in} BCI, \ie, a generic method applicable to various types of BCIs.

In addition, those stacked CNN-based methods \cite{ko2018deep, schirrmeister2017deep, gao2019eeg, emami2019seizure} have numerous trainable parameters, thus requiring large amounts of training samples, whereas BCIs generally acquire a limited number of EEG trials \cite{jayaram2016transfer}. Therefore, generalizing conventional stacked CNN-based methods in BCI is quite difficult because deep learning is a \emph{data-hungry} problem, \ie, rarely generalized with a lack of data.

Finally, interpreting a learned stacked CNN from a neurophysiologically appropriate standpoint \cite{haufe2014interpretation} is quite complicated because the CNN identifies complex patterns of data in latent space making a direct explanation difficult \cite{haufe2014interpretation}. 

In this study, we propose a novel deep learning-based BCI method to mitigate the previously discussed difficulties. The main contributions of our study are as follows:
\begin{itemize}
	\item First, we propose a novel CNN architecture, that is applicable independently from the input paradigm or type of EEG and can represent multi-scale spatio-spectral-temporal features.
	
	\item Second, the proposed method achieved positive performance on five different datasets for four differnt paradigms (two for active BCIs and two for passive BCIs). The proposed method outperformed or was similar to  state-of-the-art linear and deep learning methods, which were individually designed for each specific paradigm.
	
	\item Last, we analyze the proposed network using a variety of techniques.
\end{itemize}

The rest of this paper is organized as follows: Section \ref{sec: related work} reviews previous research on various EEG representation learning via linear model-based or deep learning-based methods. In Section \ref{sec: methods}, we propose a novel and compact deep CNN that classifies multi-paradigm EEG by representing multi-scale spatio-spectral-temporal features. Section \ref{sec: experiments} presents experimental settings and results by comparing the proposed method and comparable baselines. In Section \ref{sec: analysis and discussions}, we analyze our proposed method from several points of view. Finally, Section \ref{sec: conclusion} summarizes the proposed study and suggest future research directions.

\section{Related Work}
\label{sec: related work}
Learning a class-discriminative feature representation of EEG is still challenging in both theory and practice. Numerous prior studies have attempted to extract features from EEGs. In this section, we briefly discuss linear methods and deep learning models used for EEG signal representation.
\subsection{Linear Models}
\label{subsec: linear models}
Over the past decades, CSP \cite{blankertz2008optimizing} and its variants \cite{ang2008filter, suk2012novel} have played an essential role in decoding MI. Blankertz \etal~\cite{blankertz2008optimizing} and Ang \etal~\cite{ang2008filter} independently used a spatial filtering-based method for classifying MI. Ang \etal~\cite{ang2008filter} band-pass filtered EEG data before applying CSP, thereby attempting to decode EEG signals in a spatio-spectral manner. They named the proposed method filter bank CSP (FBCSP). Furthermore, Suk and Lee \cite{suk2012novel} also decoded MI by jointly optimizing multi spectral filters in a Bayesian framework.

CCA is commonly utilized for detecting SSVEP \cite{nakanishi2015comparison} owing to its practical ability to be implemented without the calibration stage. The standard CCA method \cite{nakanishi2015comparison} deployed sinusoidal signals as reference signals and estimated canonical correlation between the reference signals and input EEG signals to identify an evoked frequency in SSVEP EEGs.

In addition, to characterize the sleep stage, entropy calculation-based approaches were frequently used. Sanders \etal~\cite{sanders2014sleep} classified the sleep stage using the spectral-temporal features of EEGs learned from short-time Fourier transformation. Furthermore, Zheng and Lu \cite{zheng2017multimodal} focused on identifying a driver's mental fatigue during driving. They \cite{zheng2017multimodal} applied filter banks to EEG signals to extract spectral information, and then transformed the filtered EEG signals to spectral space, \ie, estimated PSD of filtered EEG signals. By doing so, Zheng and Lu \cite{zheng2017multimodal} effectively assessed the regression score of the driver's mental states which were labeled using the PERCLOS index, a measure of neurophysiological fatigue.

Earlier, Shoeb and Guttag \cite{shoeb2010application} applied a machine learning approach to extract and classify the spatio-spectral-temporal features of epileptic seizure EEG signals. Specifically, these authors \cite{shoeb2010application} used filter banks in a channel-wise manner to capture the spatio-spectral information. Then, by encoding the temporal evolution of extracted spatio-spectral feature vectors, they \cite{shoeb2010application} effectively constructed epileptic seizure EEG signal spatio-spectral-temporal features and classified the seizure and non-seizure features utilizing a support vector machine (SVM). Recently, spectral features derived from a principal component analysis (PCA) \cite{lee2017early} exhibited superior performance for seizure onset detection. In particular, Lee \etal~\cite{lee2017early} band-pass filtered raw signals and calculated PSD. Then they \cite{lee2017early} applied PCA for the extraction of EEG signal spectral features.

These practical linear model-based BCI methods \cite{blankertz2008optimizing, ang2008filter, suk2012novel, nakanishi2015comparison, lee2017early, sanders2014sleep, zheng2017multimodal} have demonstrated credible performance. However, these methods need to have certain prior neurophysiology knowledge \cite{lawhern2018eegnet}, because their feature extraction stages are specifically designed for each EEG paradigm. Conversely, our method does not need to be specialized for different paradigms.
\subsection{Deep and Hierarchical Models}
\label{subsec: deep and hierarchical models}
Recently, deep learning methods, especially CNNs have achieved promising results in EEG signal decoding researches. For instance, Schirrmeister \etal~\cite{schirrmeister2017deep} introduced Shallow ConvNet, Deep ConvNet, Hybrid ConvNet, and Residual ConvNet. These authors \cite{schirrmeister2017deep} evaluated how well various proposed CNNs decoded MI. Ko \etal~\cite{ko2018deep} also proposed a novel CNN architecture which is inspired by a recurrent convolutional neural network \cite{liang2015recurrent} for MI classification, deep recurrent spatio-temporal neural network (RSTNN).

While a standard CCA \cite{nakanishi2015comparison} has obtained state-of-the-art performance in SSVEP BCI, Kwak \etal~\cite{kwak2017convolutional} developed a CNN for SSVEP feature representation learning. These authors simply combined spatial and temporal convolution to enable the system to learn data patterns in the latent space, thereby correctly generalizing EEG signal features. Meanwhile, Waytowich \etal~\cite{waytowich2018compact} applied EEGNet \cite{lawhern2018eegnet} to the SSVEP paradigm and achieved a higher performance than that of the standard CCA \cite{nakanishi2015comparison}.

Supratak \etal~\cite{supratak2017deepsleepnet} developed a deep neural network for sleep stage detection. More precisely, they combined a CNN for representation learning and a recurrent neural network for sequential residual learning \cite{supratak2017deepsleepnet}. Furthermore, they trained the deep learning model in two separate steps, optimizing the model by individual pre-training and fine-tuning. In the meantime, Gao \etal~\cite{gao2019eeg} proposed an EEG-based spatio-temporal convolutional neural network (ESTCNN) for driver fatigue evaluation. The ESTCNN \cite{gao2019eeg} simply convolved the band-pass filtered EEG to represent temporal dependencies and flattened the extracted features for spatial features fusion. Lastly, densely connected layers were used for the identification of a user's condition \cite{gao2019eeg}.

To detect a seizure type, Asif \etal~\cite{asif2019seizurenet} proposed a multi-spectral deep feature learning using a deep CNN, SeizureNet. These authors \cite{asif2019seizurenet} transformed the EEG signals to spectral space using saliency-encoded spectrogram generation and fed the extracted spectral features to a deep neural network. In the meantime, Emami \etal~\cite{emami2019seizure} independently proposed another CNN-based approach for detecting seizure onset. They \cite{emami2019seizure} band-pass filtered and segmented the input EEG patterns, and then used a deep CNN for classification.

Recently, Lawhern \etal~\cite{lawhern2018eegnet, waytowich2018compact} proposed a novel CNN, \emph{so-called} EEGNet. Unlike other linear or deep learning-based methods, the EEGNet classified various EEG paradigms using a single architecture, \ie, not specifically tuned for different paradigms. Further, Lawhern \etal~\cite{lawhern2018eegnet} introduced a separable convolution \cite{chollet2017xception} and used it as a parameter reduction method.

On the one hand, the deep and hierarchical models decoded the EEG signals well without any custom feature extraction stage for their respective paradigm \cite{schirrmeister2017deep, ko2018deep, kwak2017convolutional, asif2019seizurenet, emami2019seizure, supratak2017deepsleepnet, gao2019eeg} or even various paradigms \cite{lawhern2018eegnet, waytowich2018compact}. On the other hand, the deep CNNs extracted the EEG features at a sequential level using stacked convolutional layers without exploiting multi-scale spectral representation. Conversely, the proposed method exploits multi-scale spatio-spectral-temporal features irrespective of the input EEG paradigms.


\begin{figure}[t]\centering
	\includegraphics[width=.85\linewidth]{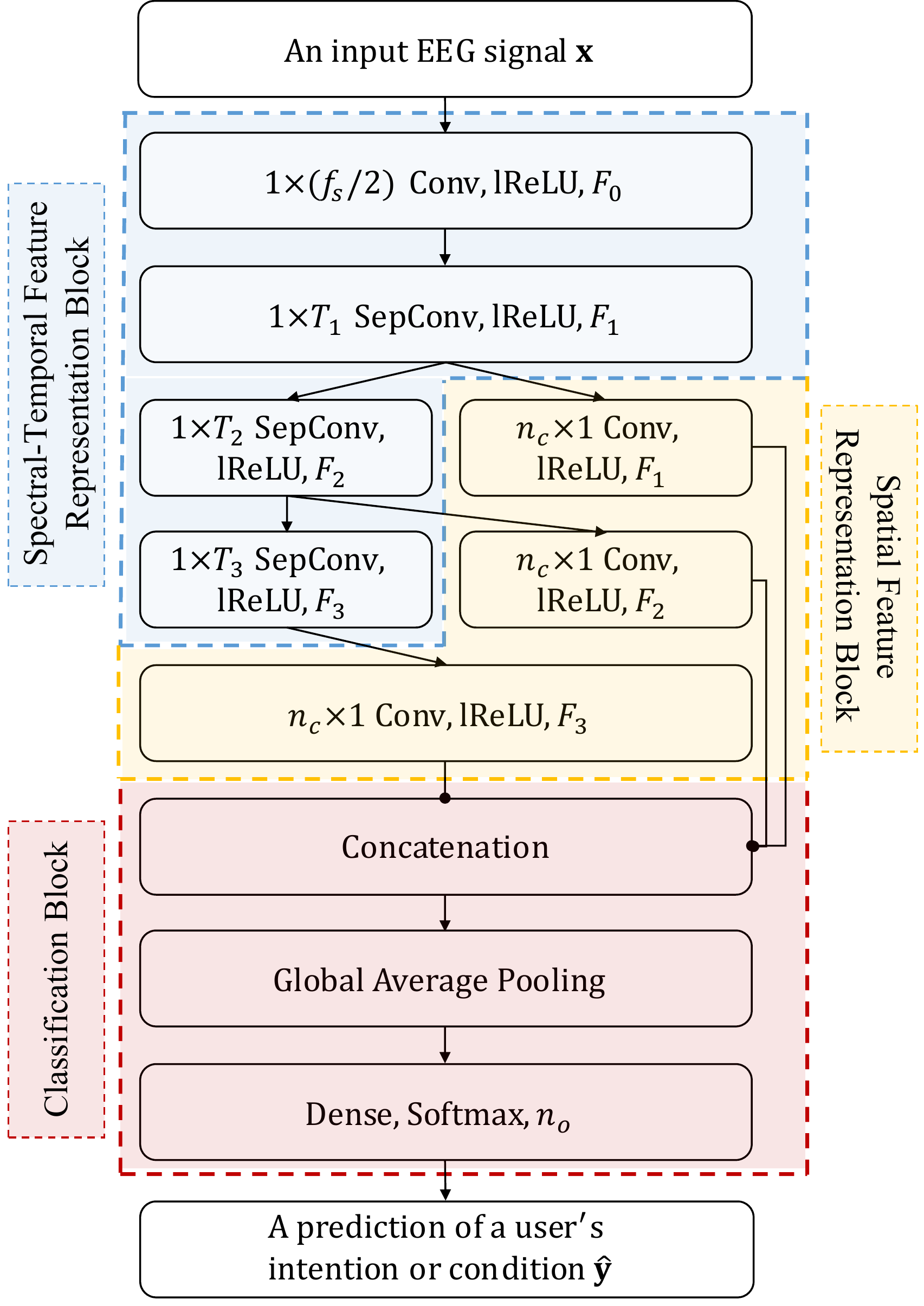}
	\caption{Architectural framework of our multi-scale neural network (MSNN). In the proposed network, first, an input EEG $\mathbf{x}$ is temporally convolved to expand the number of features, where $f_s$, $F_0$, lReLU denote the sampling frequency rate, the number of output filter maps of the first layer, and leaky rectified linear unit activation function respectively. Then, a set of temporal separable convolutions extracts \emph{spectral-temporal} features ($T_k$ and $F_k$ respectively denote the kernel size and output feature maps of $k$-th temporal separable convolution). At the same time, a set of spatial convolutions represents \emph{spatial} features, where $n_c$ denotes the number of acquired EEG channels. Then, the \emph{multi-scale} features are concatenated and fed into the global average pooling layer \cite{lin2013network}. Finally, the dense layer classifies the class of input EEG by exploiting multi-scale features where $n_o$ denotes the number of output nodes.}
	\label{fig: msnn_architecture}
\end{figure}

\section{Methods}
\label{sec: methods}
In this section, we propose a deep multi-scale neural network (MSNN), which can represent EEG features from different paradigms by exploiting \emph{spatio-spectral-temporal} information at multi-scale.

\subsection{Multi-Scale Neural Network}
\label{subsec: multi-scale neural network}
As mentioned previously, an FBCSP \cite{ang2008filter} is one of the most successful models to exploit EEG signal \emph{multi-scale} features, especially for MI. Thus, many successful MI EEG signal decoding algorithms \cite{schirrmeister2017deep, suk2012novel} or even other paradigm classification algorithms \cite{lawhern2018eegnet} are inspired by the FBCSP \cite{ang2008filter} model. In this study, the proposed multi-scale neural network (MSNN) also learns multi-scale feature representations. However, the network automatically learns from data through discriminative multiple spectral filters, rather than manually defining multi-frequency bounds as in FBCSP \cite{ang2008filter}. Basically, our proposed method consists of three types of blocks: (1) a \emph{spectral-temporal} feature representation block, (2) a \emph{spatial} feature representation block, and, (3) a classification block, as depicted in Fig \ref{fig: msnn_architecture}.

First, in the spectral-temporal feature representation block, stacked convolutional layers extract EEG data spectral-temporal features, such as existing EEG classification methods. However, the proposed model exploits intermediate activations for gathering multi-scale spectral information. Then, the spatial feature representation block discovers spatial patterns from the extracted multi-scale features. Finally, these multi-scale spatio-spectral-temporal features are concatenated, pooled, and fed into the densely connected layer for classification.


\subsection{Spectral-Temporal Feature Representation Block}
\label{subsec: spectral-temporal feature representation blocks}
Given an input EEG data $\mathbf{x}$, we reshape it in the form of $[n_c, n_T, 1]$, \ie, $x\in\mathbb{R}^{n_c\times n_t\times 1}$, where $n_c$ and $n_T$ denote the number of electrode channels and timepoints, respectively.

In the MSNN, the input EEG data are temporally convolved in a channel-wise manner by a temporal convolutional layer to expand the number of feature maps. Thus, the activated features have the form $[n_c, n_T', F_0]$, where $n_T'=n_T-(f_s/2)+1$ ($f_s$ and $F_0$ are the sampling frequency and the feature map dimension for the first temporal convolution layer.). The main benefits of using a separable convolution \cite{chollet2017xception, lawhern2018eegnet} are a significant reduction of tunable weights in the model and, more importantly, an efficient and explicit decoupling of the relationship between the temporal and the feature map dimensions of the input features. This is accomplished by learning kernels independently for each feature map. Thus, as in BCI literature, the separable convolution \cite{chollet2017xception} enables the system to learn temporal kernels individually from the feature map dimensions (using a depthwise convolution \cite{chollet2017xception}), and then optimally re-combine the feature maps (using a pointwise convolution \cite{chollet2017xception}).

In this block, by setting a kernel size of $(1\times T_k)$, where $T_k$ denotes the kernel size of the $k$-th temporal separable convolution, the $k$-th temporal separable convolutional layer represents EEG signal features in the range of $T_k/f_s'$ sec, hence, $f_s'/T_k$ Hz, where $f_s'$ is a frequency property extracted at the first temporal convolutional layer. Therefore, the spectral-temporal feature representation layers can deal with different timepoints or frequency ranges by using various kernel sizes for the input EEG data.

Additionally, each different layer that has a different kernel size extracts features in different frequency and timepoint ranges. In other words, a spectral-temporal convolution layer with a larger kernel represents \emph{longer-term} temporal features, \ie, a \emph{lower-range} of spectral features and vice versa. Then, the MSNN exploits intermediate activations from each layer, thus learning multi-scale feature representations.

In addition, a separable convolution \cite{chollet2017xception} only operates convolutions in a cross-$n_T$ way, thus, the number of parameters is small compared to a conventional convolution. For instance, while a $k$-th separable temporal convolution has only $T_k + F_{k-1}\cdot F_k$ parameters, the conventional convolution with the same size kernel has $T_k\cdot F_{k-1} \cdot F_k$ parameters, where $F_k$ denotes the feature maps dimension of $k$-th layer.

Furthermore, in this processing, as described above, the MSNN uses its intermediate activations to exploit multi-scale representations. In other words, the proposed network obtains $N$ numbers of spectral-temporal features $\mathbf{f}_k^\text{ST}$, $k=1, 2, \cdots, N$ like:
\begin{equation}
	\mathbf{f}_k^\text{ST} = \mathcal{C}_k^\text{ST}(\mathbf{x})= \mathcal{C}_k^\text{sep}\circ\mathcal{C}_{k-1}^\text{sep}\circ\cdots\circ\mathcal{C}_0(\mathbf{x}),
	\label{eq: 1}
\end{equation}
where $\mathcal{C}_k^\text{sep}$, $\mathcal{C}_0$, and $\mathcal{F}_i\circ\mathcal{F}_j$, respectively, denote the $k$-th separable convolution, the first temporal convolution, and a function composition between arbitrary functions $\mathcal{F}_i$ and $\mathcal{F}_j$, \ie, $\mathcal{F}_i\circ\mathcal{F}_j(\cdot)=\mathcal{F}_i(\mathcal{F}_j(\cdot))$. Thus, by extracting features $\mathbf{f}_1^\text{ST},\mathbf{f}_2^\text{ST},\cdots,\mathbf{f}_N^\text{ST}$, the MSNN effectively represents the spectral-temporal features from the multi-scale viewpoint, thereby automatically enhancing generalization. In addition, as all inputs are \emph{zero-padded} before each separable temporal convolution, the output features have the same dimension for the channels and timepoints, except for the feature map dimension. Thus, the $k$-th spectral-temporal feature $\mathbf{f}_k^\text{ST}$ now has the form $[n_c, n_T', F_k]$.
\subsection{Spatial Feature Representation Block}
\label{subsec: spatial feature representation blocks}
In the spatial feature representation block, a common spatial convolution is used for feature extraction. In this block, the kernel size is constrained to be equal to the number of EEG channels, hence, a convolution with a kernel of $(n_c\times 1)$ is used. Additionally, by setting the kernel size to be the same as the number of electrode channels, similar to many existing deep learning-based BCI methods \cite{schirrmeister2017deep, ko2018deep, lawhern2018eegnet}, the proposed MSNN extracts spatial information from the original EEG acquisition channel distributions of multi-scale spectral temporal features. Then, the MSNN can obtain \emph{neurophysiologically plausible} information from the input data distribution.

Furthermore, the spatial feature representation can be applied unrestrictedly, thus in the proposed method, we add this block after every extracted spectral-temporal features $\mathbf{f}_k^\text{ST}$, $k=1, 2, \cdots, N$ like,
\begin{equation}
	\mathbf{f}_k^{\text{SST}} = \mathcal{S}_k(\mathbf{f}_k^{\text{ST}}) = \mathcal{S}_k\circ\mathcal{C}_k^\text{ST}(\mathbf{x}),
	\label{eq: 2}
\end{equation}
where $\mathcal{S}_k$ denotes the $k$-th spatial convolution and $\mathbf{f}_k^\text{SST}$ is \emph{spatio-spectral-temporal} features estimated by the $\mathcal{S}_k$ and $\mathcal{C}_k^\text{ST}$. We use \emph{valid paddings} for every spatial convolution, thus the $k$-th spatio-spectral-temporal feature $\mathbf{f}_k^\text{SST}$ has the form $[1, n_T', F_k]$. By setting the number of spatial convolutions to be identical to the number of spectral-temporal convolutions, unlike many previous researches using deep learning for BCI \cite{schirrmeister2017deep, ko2018deep, lawhern2018eegnet, asif2019seizurenet, supratak2017deepsleepnet}, we extract spatial features of each range from spectral-temporal features. In other words, unlike many previous stacked CNNs, the proposed architecture uses every intermediate activated feature set to exploit spatial information, thereby creating the capability to extract various ranges of EEG features at multi-scale.
\subsection{Classification Block}
\label{subsec: classifier learning}
For classifier learning, because we have $N$ numbers of different (or same when $F_1=F_2=\cdots=F_N$) size of spatio-spectral-temporal features $\mathbf{f}_k^\text{SST}$, $k=1,2,\cdots,N$, the classifier in the proposed method has to concatenate the features in the feature map dimension. Thus, the concatenated feature $\mathbf{f}_\text{concat}^\text{SST}$ is represented as:
\begin{equation}
	\mathbf{f}_\text{concat}^\text{SST}=\concat_{i=1}^N \mathbf{f}_i^\text{SST}=\concat_{i=1}^N \mathcal{S}_i \circ \mathcal{C}_i^\text{ST}(\mathbf{x}),
	\label{eq: 3}
\end{equation}
where $\concat$ denotes the concatenation operation.

For the classifier network, let us assume that the number of output classification nodes is denoted by $n_o$ and we use a single linear mapping layer. Then, we need to train the large number of $\sum_{i=1}^{N}n_o\cdot n_T' \cdot F_i$ parameters (note that we disregard the bias term for a convenient calculation) because $\mathbf{f}_\text{concat}^\text{SST}$ has the form $[1, n_T', \sum_{i=1}^{N}F_i]$, and it would still require a large number of training samples. Therefore, after representing the input EEG data multi-scale spatio-spectral-temporal features, the proposed MSNN has one extra operation for reducing the trainable weights. Unlike the existing deep learning-based BCI methods \cite{lawhern2018eegnet, schirrmeister2017deep, ko2018deep, kwak2017convolutional, asif2019seizurenet, supratak2017deepsleepnet}, global average pooling (GAP), which is widely used in the computer vision field \cite{lin2013network} is performed.

The GAP layer \cite{lin2013network}, a type of pooling layer, averages nodes from each feature map, thus eliminating the requirement for any window size or stride. By applying GAP \cite{lin2013network}, our proposed MSNN efficiently extracts significant features. From the BCI literature, the GAP layer \cite{lin2013network} can be understood to be a method that can emphasize an important frequency range and its surrounding area for each feature map dimension. Thus, for the extracted multi-scale features in the MSNN, the GAP layer \cite{lin2013network} stresses the crucial spectral-temporal part resulting in concise information for the final decision making.

Additionally, the GAP layer \cite{lin2013network} significantly reduces the number of classifier parameters used in the proposed MSNN. Specifically, after the GAP layer $\mathcal{G}(\cdot)$, the extracted feature is reduced to the form $[1, 1, \sum_{i=1}^{N}F_i]$, whereas the feature without GAP has the form $[1, n_T', \sum_{i=1}^{N}F_i]$. Therefore, we drastically suppress the trainable parameters in the classifier from $n_T'\cdot n_o\cdot\sum_{i=1}^N F_i$ to $n_o\cdot\sum_{i=1}^N F_i$.

Then, the MSNN prediction, $\hat{\mathbf{y}}$, for the input EEG data, $\mathbf{x}$, is as follows:
\begin{align}
\hat{\mathbf{y}} &= \mathrm{softmax}\left(\mathbf{W}_o^\top\cdot\mathcal{G}\left(\mathbf{f}_\text{concat}^\text{SST}\right)+\mathbf{b}_o\right) \nonumber\\
&= \mathrm{softmax}\left(\mathbf{W}_o^\top\cdot\mathcal{G}\left[\concat_{i=1}^N \mathcal{S}_i\circ\mathcal{C}_i^\text{ST}(\mathbf{x})\right] + \mathbf{b}_o\right),
\label{eq: 4}
\end{align}
where $\mathbf{W}_o\in\mathbb{R}^{\sum_{i=1}^N F_i\times n_0}$ and $\mathbf{b}_0\in\mathbb{R}^{n_o}$ respectively denote the weight matrix and bias of the classifier. 

Finally, the cross-entropy loss, $\mathcal{L}$, that is used for network training is calculated by the prediction $\hat{\mathbf{y}}$ and the label $\mathbf{y}$:
\begin{equation}
\mathcal{L}=\mathrm{CE}(\mathbf{y},\hat{\mathbf{y}})=-\sum_{b=1}^{B} \mathbf{y}^{(b)}\log \hat{\mathbf{y}}^{(b)},
\label{eq: 5}
\end{equation}
where $B$ and $\mathrm{CE}(\cdot,\cdot)$ respectively denote the mini-batch sizes and the cross-entropy loss function, and $\hat{\mathbf{y}}^{(b)}$ and $\mathbf{y}^{(b)}$ denote the prediction and ground-truth label for the $b$-th training sample in the mini-batch\footnote{All codes used in our experiments are available at `\url{https://github.com/DeepBCI/Deep-BCI/tree/master/1_Intelligent_BCI/Multi_Scale_Neural_Network_for_EEG_Representation_Learning_in_BCI}.'}.
\section{Experiments}
\label{sec: experiments}
In this section, we describe the datasets used for performance evaluation, our experimental settings, and baseline settings. Furthermore, we present the performance of our method and competing methods.

\subsection{Datasets and Preprocessing}
\label{subsec: datasets and preprocessing}
In this study, we used five different publicly available datasets to validate the proposed method on four different EEG data paradigms.
\subsubsection{Motor Imagery}
First, we used two big datasets for MI EEGs, GIST-MI \cite{cho2017eeg}\footnote{Available at \url{http://gigadb.org/dataset/100295}} and KU-MI \cite{lee2019eeg}\footnote{\label{note1}Available at \url{http://gigadb.org/dataset/100542}}\textsuperscript{,}\footnote{Experimental results of the KU-MI dataset \cite{lee2019eeg} are reported in Supplementary B.}. The GIST-MI \cite{cho2017eeg} dataset consists of two different MI tasks: left-hand and right-hand MI that are acquired from 52 subjects. All EEG signals were recorded from 64 Ag/AgCl electrode channels according to the standard 10-20 system, sampled at 512Hz. Each class contained 100 or 120 trials, and each trial was a 3 sec long MI task. Because this dataset is not separated into training and test samples, we conducted a five-fold cross-validation for a fair evaluation. For the MI datasets, we preprocessed signals by applying a large Laplacian filtering\footnote{When the target channel does not have four nearest neighbors, we just used available channels and their average value to filter the target channel.}, baseline correction by subtracting the mean value of the fixation signal from each MI trial, and band-pass filtering between 4 and 40Hz. Then, we removed the first and last 0.5 sec from each trial, and finally applied Gaussian normalization. We applied the same mean and standard deviation values for normalization to the test samples. The multi-channel EEG signals were only shifted and scaled by their respective channel-wise mean and standard deviation values. Thus, inter-channel relations inherent in the data were preserved.

\subsubsection{Steady-State Visually Evoked Potentials}
We also used the KU-SSVEP dataset \cite{lee2019eeg}\textsuperscript{\ref{note1}} for SSVEP decoding experiments in this study. This KU-SSVEP dataset \cite{lee2019eeg} was acquired from 54 subjects and recorded from 62 Ag/AgCl electrode channels using the 10-20 system. The KU-SSVEP dataset \cite{lee2019eeg} contains four EEG classes from target stimuli at 5.45, 6.67, 8.57, and 12Hz, and each class has 25 EEG trials of training and testing samples for each session. We preprocessed the SSVEP signals by applying band-pass filtering between 4 and 15Hz and selected eight channels in the occipital region, `PO3, POz, PO4, PO9, O1, Oz, O2, and PO10,' because this region is widely used for SSVEP classification \cite{waytowich2018compact}.

\subsubsection{Drowsiness}
With respect to passive BCI \cite{arico2018passive}, we considered two different paradigms, seizure EEG signals \cite{shoeb2009application} and vigilance EEG signals \cite{zheng2017multimodal}. Owing to its theoretical and practical benefits, in this study, we conducted experiments identifying drivers' mental fatigue. We also used a publicly available SEED-VIG EEG dataset \cite{zheng2017multimodal}\footnote{Available at: \url{http://bcmi.sjtu.edu.cn/seed/download.html}} for the drowsy driving task data. This dataset \cite{zheng2017multimodal} consists of 23 experiments, \ie, trials, and each trial is recorded for approximately 2 hours while simulated driving occurs. The EEG signals are acquired from 17 electrode channels according to the 10-20 system and sampled at 200Hz \cite{zheng2017multimodal}.  For this dataset, we band-pass filtered EEG signals in the range between 0.5 and 40Hz, each epoch was 8 sec in length. Because the dataset was originally labeled using \emph{PERCLOS} levels \cite{zheng2017multimodal}, we categorized the label vectors into three classes, \emph{awake}, \emph{tired}, and \emph{drowsy} with two threshold  values(0.35 and 0.7) \cite{zheng2017multimodal}. Then, for every 23 experiments, a five-fold cross-validation was used for performance estimations.

\subsubsection{Seizure}
Finally, we conducted seizure onset detection experiments with the widely used and publicly available CHB-MIT \cite{shoeb2009application}\footnote{Avaliable at: \url{https://physionet.org/content/chbmit/1.0.0/}} dataset. The CHB-MIT dataset \cite{shoeb2009application} contains EEG data from 24 subjects sampled at 256Hz acquired from 23 electrode channels (24 or 26 in a few cases) according to the 10-20 system. In this work, we selected EEG trials that have the same 23 channels montage and removed some trials acquired from the different montage. By following \cite{shoeb2010application}, we used a \emph{leave-one-record-out} cross-validation. More precisely, we trained the proposed method using all non-seizure records and all seizure records but one, and tested the model on the remaining seizure record \cite{shoeb2010application}. Then, we repeated this process for the number of seizure records in the dataset, thus, each seizure record was tested. For training, the test trial epochs were 10 sec in length. During validation and testing session, a 10 sec length EEG signal was input into the proposed network using a 1/256 stride. Then, we observed whether the probability values for each EEG signal timepoint was ictal or normal.

For all datasets, the training samples were randomly selected and split again into training and validation samples for model selection. Specifically, we divided the training samples at a 9:1 ratio for each subject and used them for training and model selection respectively.

\begin{table*}[t]
	\caption{\label{table: I}Performance evaluations. The Method column denotes all used classification/detection methods including baselines and the proposed method on the various datasets, GIST-MI \cite{cho2017eeg}, KU-SSVEP \cite{lee2019eeg}, SEED-VIG \cite{zheng2017multimodal}, and CHB-MIT \cite{shoeb2009application} EEG dataset. Each cell depicts the average performance and the standard deviation of all subjects (or trials for the SEED-VIG \cite{zheng2017multimodal}). For classification performance on the SSVEP dataset, we used different kernel sizes for EEGNet \cite{waytowich2018compact} and the proposed method. These values are marked by $\dagger$ and $\ddagger$, respectively.}
	\centering
	{\renewcommand{\arraystretch}{1.1}
		\begin{tabular}{|c||c|c|cc|c|}\hline
			\multirow{3}{*}{Method} & GIST-MI \cite{cho2017eeg}  & KU-SSVEP \cite{lee2019eeg}  & \multicolumn{2}{c|}{SEED-VIG \cite{zheng2017multimodal}}& CHB-MIT \cite{shoeb2009application}\\\cline{2-6}
			& \multicolumn{2}{c|}{Classification accuracy} & \multicolumn{3}{c|}{Number of false detections} \\\cline{2-6}
			& \multicolumn{2}{c|}{Mean$\pm$Std.}  & Mean$\pm$Std. & False Positive (Drowsy) & Mean (Mean latency)\\\hline
			CSP + LDA \cite{blankertz2008optimizing}	&	.66$\pm$.14		&	-			&-	&- &-	\\
			FBCSP + LDA \cite{ang2008filter}			&	.68$\pm$.15		&	-			&-	&- &-	\\
			CCA \cite{nakanishi2015comparison}			&	-						&	.94$\pm$.09	&-	&- &-	\\
			PSD + SVM \cite{zheng2017multimodal}			&		-			&-		&31.20$\pm$15.47 & 6.74	&-	\\
			Shoeb and Guttag \cite{shoeb2010application}				&		-		&	-		&- & -& 5.35 (5.11)	\\
			\hline
			Shallow ConvNet \cite{schirrmeister2017deep}&	.63$\pm$.11		&	.52$\pm$.20		&34.89$\pm$19.13 & 6.51& 19.21 (8.48)	\\
			Deep ConvNet \cite{schirrmeister2017deep}	&	.61$\pm$.07	&\bf.96$\pm$.08		& 41.31$\pm$21.04 & 8.65& 8.74 (7.52)	\\
			RSTNN \cite{ko2018deep}	&	.69$\pm$.12	&	.65$\pm$.20		&39.84$\pm$22.56 & 8.08&24.35 (9.31)	\\
			ESTCNN \cite{gao2019eeg}	&.67$\pm$.10		&	.79$\pm$.17		&41.10$\pm$21.31& 8.71& 6.41 (7.01)	\\
			EEGNet \cite{lawhern2018eegnet, waytowich2018compact}				&	.64$\pm$.07		&	.93$\pm$.10$^\dagger$			& 46.63$\pm$22.10&11.26& 5.40 (6.23)	\\
			\hline
			MSNN (Proposed)									&\bf.81$\pm$.12		&	.93$\pm$.08$^\ddagger$	&\bf 31.10$\pm$17.29 &\bf 5.38	&\bf 5.35 (4.98)	\\\hline
	\end{tabular}}
\end{table*}

\subsection{Experimental Settings}
\label{subsec: experimental settings}
In our work, we compared our method with paradigm-specific linear model-based and deep learning-based methods for each EEG paradigm. 
\subsubsection{Linear Models - Motor Imagery}
First, we built a CSP with a linear discriminant analysis (CSP + LDA) \cite{blankertz2008optimizing} and an FBCSP with an LDA (FBCSP + LDA) \cite{ang2008filter} for MI decoding. We used four filters and regularized covariance for the CSP \cite{blankertz2008optimizing} and FBCSP \cite{ang2008filter}. Additionally, we also used nine non-overlapped filter banks in the 4$\sim$40Hz range, \ie, 4$\sim$8, 8$\sim$12, $\cdots$, 36$\sim$40Hz, and, finally selected 10 features using the mutual information-based feature selection method FBCSP \cite{ang2008filter}. 

\subsubsection{Linear Models - Steady-State Visually Evoked Potentials}
We also built a standard CCA \cite{nakanishi2015comparison} for SSVEP classification. We set reference signals for each stimulus including second harmonics. Furthermore, the standard CCA \cite{nakanishi2015comparison} does not require training samples for the optimization, thus we only estimated each session in its entirety from the KU-SSVEP dataset \cite{lee2019eeg} for the CCA performance estimation. 

\subsubsection{Linear Models - Drowsiness}
For the drowsy state detection experiment, we estimated the filter-banked input EEG data PSD in a channel-wise manner for extracting spatio-spectral features and classified the learned features using an SVM with a radial basis function (RBF) kernel ($\gamma=1/d_\text{input}$ where $d_\text{input}$ denotes the input feature dimension) \cite{zheng2017multimodal}.

\subsubsection{Linear Models - Seizure}
In addition, we also reimplemented Shoeb and Guttag \cite{shoeb2010application}'s method for the seizure onset detection experiment. We applied the PSD to the EEG data in a channel-wise manner. Then, the 3 sec time window time evolution \cite{shoeb2010application} method was used for capturing temporal information. Finally, the represented spatio-spectral-temporal features were fed into an SVM using an RBF kernel ($\gamma=1/d_\text{input}$). 
\subsubsection{Deep Neural Networks - Motor Imagery}
We also implemented deep learning-based BCI models\footnote{See `Appendix A: Architectural Details of Deep Models for BCIs' for more detail architectures and learning schedules.} for MI. Basically, most of the existing deep learning models \cite{schirrmeister2017deep, ko2018deep, emami2019seizure, gao2019eeg} have focused on a paradigm-specific BCI task. However, we conducted experiments over all types of datasets for each deep learning model to demonstrate the validity of the proposed method. We built a Shallow ConvNet and a Deep ConvNet as proposed by Schirrmeister \etal~\cite{schirrmeister2017deep}. The Shallow ConvNet \cite{schirrmeister2017deep} consists of two convolutions, temporal and spatial, with a squaring nonlinear activation, an average pooling, and a logarithmic activation. The Deep ConvNet \cite{schirrmeister2017deep} has five convolutions, temporal and spatial, and three additional temporal convolutions. The RSTNN \cite{ko2018deep} is also used for these experiments. This network \cite{ko2018deep} consists of three recurrent convolutional layers, and each recurrent convolutional layer has three recurrent temporal convolutions \cite{liang2015recurrent} with a spatial convolution. 

\subsubsection{Deep Neural Networks - Steady-State Visually Evoked Potentials}
For the SSVEP decoding experiment, we exploited another version of EEGNet for SSVEP EEG \cite{waytowich2018compact}. We used different kernel sizes for this EEGNet \cite{waytowich2018compact} as Waytowich \etal~proposed. The SSVEP classification performance estimated by this version \cite{waytowich2018compact} is marked by $\dagger$ in the classification table.

\subsubsection{Deep Neural Networks - Drowsiness}
The ESTCNN \cite{gao2019eeg} which is proposed for mental fatigue classification has three core blocks. Each block in the ESTCNN \cite{gao2019eeg} consists of three temporal convolutions with a max pooling layer with the exception of the last block that uses an average pooling layer instead of the max pooling. 

\subsubsection{Deep Neural Networks - Multi-paradigm}
Finally, we also implemented the EEGNet \cite{lawhern2018eegnet} in our study. As previously mentioned, we used different kernel sizes for two different EEGNets, \cite{lawhern2018eegnet} and \cite{waytowich2018compact}. Nevertheless, the basic architecture of the network was the same for various EEG paradigms, having a temporal convolution, depthwise spatial convolution \cite{chollet2017xception}, and separable temporal convolution \cite{chollet2017xception}.

\subsubsection{Proposed Multi-Scale Neural Network}
\label{subsubsec: msnn}
While training our proposed network, depicted in Fig. \ref{fig: msnn_architecture}, we set a mini-batch size of 16, an exponentially decreasing learning rate (initial value: 0.03, decreasing ratio per epoch: 0.001), and an Adam optimizer. For the first temporal convolution, we used a conventional temporal convolution with the kernel size of ($1\times f_s/2$) and $F_0=4$. Furthermore, we used three spectral-temporal feature representation convolutions, \ie, $N=3$, and set $T_1=100$, $T_2=60$, and $T_3=20$ with $F_1=16$, $F_2=32$, and $F_3=64$. Then, for the spatial feature representation block, we used three spatial convolutions because the number of spatial convolutional layers must be the same as the number of spectral-temporal separable convolutional layers. The proposed method used different kernel sizes for the SSVEP dataset, similar to the EEGNet \cite{waytowich2018compact} due to the fact that SSVEP EEG data is created by target frequencies \cite{lee2019eeg, nakanishi2015comparison}. For the KU-SSVEP dataset \cite{lee2019eeg}, we set $T_1=20$, $T_2=10$, and $T_3=5$ for the spectral-temporal feature representation block, and used the same settings for the others. The SSVEP classification performance estimated by this method is marked by $\ddagger$. Additionally, batch normalization was performed after every convolution. Finally, for the classification block, all activated features from the \emph{spatio-spectral-temporal} block were concatenated and fed into the GAP \cite{lin2013network} layer. Then, after flattening, the \emph{multi-scale} features were linearly mapped by a dense layer. In this proposed network, a leaky rectified linear unit (ReLU) activation function, an L1-L2 regularizer ($\ell_1=0.01$ and $\ell_2=0.001$), and a Xavier initializer \cite{glorot2010understanding} are used for all tunable parameters except for the final decision layer that is activated by a softmax activation function instead of a leaky ReLU. We selected model components that demonstrated the best performance for validation, \ie, model selection samples, as mentioned previously.

\subsection{Experimental Results}
\label{subsec: experimental results}
\subsubsection{Motor Imagery}
All experimental results are summarized in TABLE \ref{table: I}. Our proposed network clearly outperformed other baselines for MI EEG signal decoding. Importantly, the proposed network achieved a higher accuracy than those methods designed specifically for MI classification: CSP \cite{blankertz2008optimizing}, FBCSP \cite{ang2008filter}, Shallow ConvNet \cite{schirrmeister2017deep}, Deep ConvNet \cite{schirrmeister2017deep}, and RSTNN \cite{ko2018deep}. With this clear improvement in accuracy, we could expect that our proposed method is one step closer to MI-based BCI commercialization.

\subsubsection{Steady-State Visually Evoked Potentials}
Our proposed MSNN achieved a slightly lower performance than CCA \cite{nakanishi2015comparison}, Deep ConvNet \cite{schirrmeister2017deep}, and EEGNet \cite{waytowich2018compact} in the SSVEP classification. However, the difference in performance between our MSNN and the other three baselines, CCA \cite{nakanishi2015comparison}, Deep ConvNet \cite{schirrmeister2017deep}, and EEGNet \cite{waytowich2018compact}, was reasonably small and the proposed method performed with a credible accuracy score.

\subsubsection{Drowsiness}
The proposed MSNN made the smallest number of mistakes in decision making for passive BCI \cite{arico2018passive}. In particular, the proposed method detected a driver's mental fatigue, \ie, drowsiness, from the EEG signals. Our proposed method predicted 31.10 incorrect trials from a total of 177 samples on average. Furthermore, accurately detecting a drowsy state is one of the most important MSNN capabilities for practical use. Our proposed model only made 5.38 mistakes out of 35 drowsy trials on average, thus exhibiting the highest precision score.

\subsubsection{Seizure}
Finally, the MSNN incorrectly identified 5.35 seizures among 178 total test seizure samples. Furthermore, our proposed network was the fastest for detecting seizures, \ie, it exhibited the shortest latency time (approximately 4.98 sec on average) among various methods. In other words, our proposed method demonstrated the best performance even with the shortest latency time. Additionally, the proposed model correctly identified approximately 92\% of the seizures within 4.98 sec. We do not present the standard deviation values for this seizure detection experiment because each test trial consisted of different numbers of seizures.

\section{Analyses and Discussions}
\label{sec: analysis and discussions}
In this section, we analyzed our proposed network. We determined the feature response by estimating PSD values and relevance scores \cite{montavon2017explaining} to show the multi-scale learning benefits. We also visualized learned weights and represented features of the proposed method using different methodologies, activation pattern maps \cite{haufe2014interpretation} and t-SNE plots. Additionally, we observed a practical use for the proposed method, especially for drowsiness and seizure detection experiments.

\begin{figure}[t]
	\centering
	\includegraphics[width=1\linewidth]{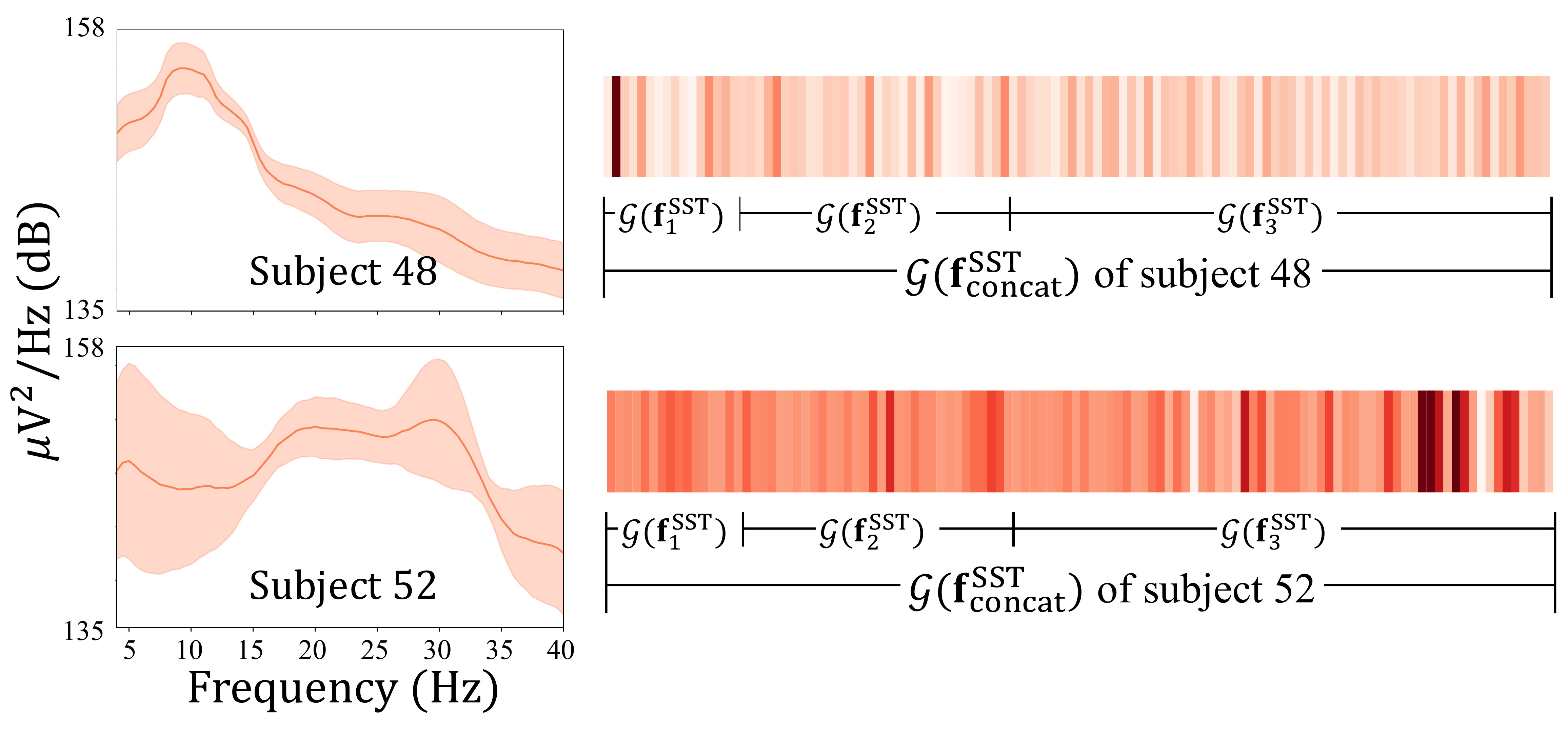}
	\caption{PSD curves (left) and relevance scores \cite{montavon2017explaining} (right) for subject 48 (top) and subject 52 (bottom) from the GIST-MI dataset \cite{cho2017eeg}. For the PSD curves, the solid red line and the shaded region exhibit the mean and standard deviation of PSD values of all trials, respectively. We observed that our proposed MSNN concentrates features from the lower frequency range for subject 48 and a wide range for subject 52.}
	\label{fig: noise_curve}
\end{figure}
\subsection{Multi-Scale EEG Feature Extraction}
To demonstrate the multi-scale information capture ability of our proposed method, we estimated and plotted PSD values and relevance scores \cite{montavon2017explaining} for MI EEG samples. Specifically, we estimated PSD values for subject 48 and 52 in the GIST-MI dataset \cite{cho2017eeg}'s EEG samples from channels on the motor cortex. Additionally, we calculated relevance scores for those subjects by a \emph{layer-wise relevance propagation} \cite{montavon2017explaining}. In our results, all classification methods evenly demonstrated well-generalization (baselines: $\sim$80\% and proposed: $\sim$85\%) for subject 48, whereas only the proposed method achieved superior performance for subject 52 (baselines: $<$65\% and proposed: $\sim$80\%). As Fig. \ref{fig: noise_curve} shows, subject 48's EEG samples are highly activated at the $\mu$ range, while subject 52's samples do not show any clear trend at the $\mu$ range, but in a wider range. Our proposed network exhibited a high relevance score at the low-frequency range for subject 48 who exhibited a clear trend at the low-frequency range. Furthermore, the relevance scores for subject 52 were roughly alike for the wider range, where subject 52's PSD demonstrated a less clearly defined trend.

From this phenomenon, we can conclude that our proposed MSNN can capture important features on the multi-scale range, not only in the frequency of interest. In other words, while other existing methods gather spatio-spectral-temporal information at the sequential level, the proposed network exploit multi-scale features, thereby improving learning ability\footnote{Randomly selected additional results are reported in Supplementary C.}.

\begin{figure*}[tbph]
	\begin{subfigure}{1\linewidth}
		\centering
		\includegraphics[width=.92\linewidth]{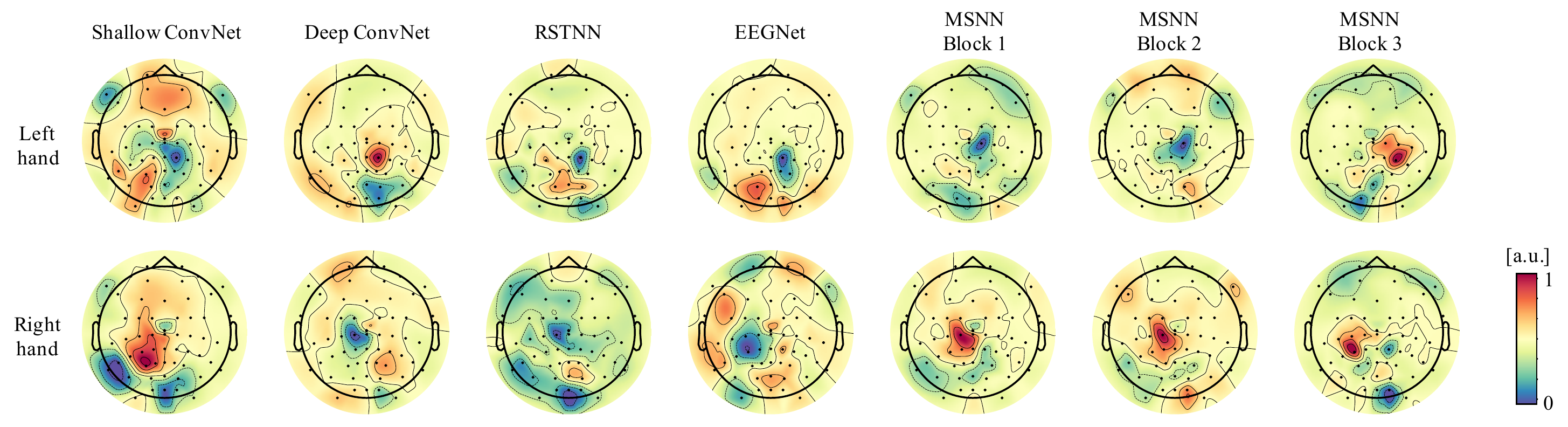}
		\caption{Topologically visualized activation pattern maps \cite{haufe2014interpretation} of comparable baselines, and three spatial convolutions in the proposed network. All these visualized patterns here are estimated by the first subject's first fold EEG signals in the GIST-MI dataset \cite{cho2017eeg} and normalized in a range between 0 and 1. Finally, [a.u.] denotes an arbitrary unit.}
		\label{subfig: activation_patterns}
	\end{subfigure}\vspace{5pt}
	\begin{subfigure}{1\linewidth}
		\centering
		\includegraphics[width=.9\linewidth]{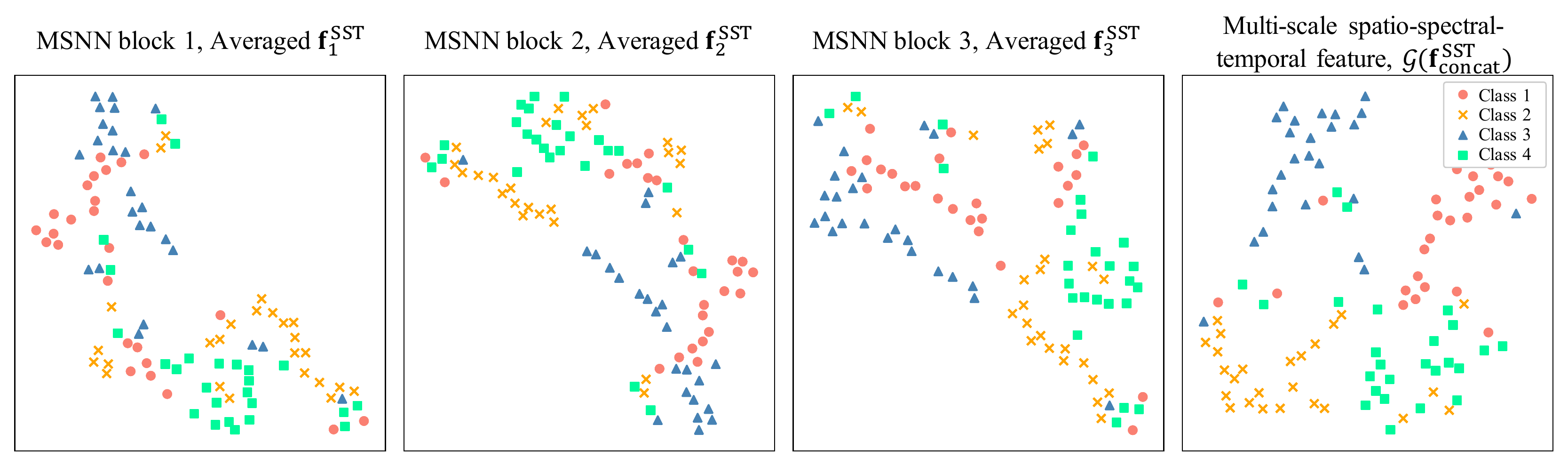}
		\caption{Visualization of t-SNE transformed represented features for test SSVEP EEG samples. The first three figures denote extracted features by the first, second, and final spatial convolutional layers of the proposed method. The final figure exhibits the GAP \cite{lin2013network}-ed feature, $\mathcal{G}(\mathbf{f}_\text{concat}^\text{SST})$ which is used for final decision making.}
		\label{subfig: tsne_plotts}
	\end{subfigure}\vspace{5pt}
	\begin{subfigure}{1\linewidth}
		\centering
		\includegraphics[width=.163\linewidth]{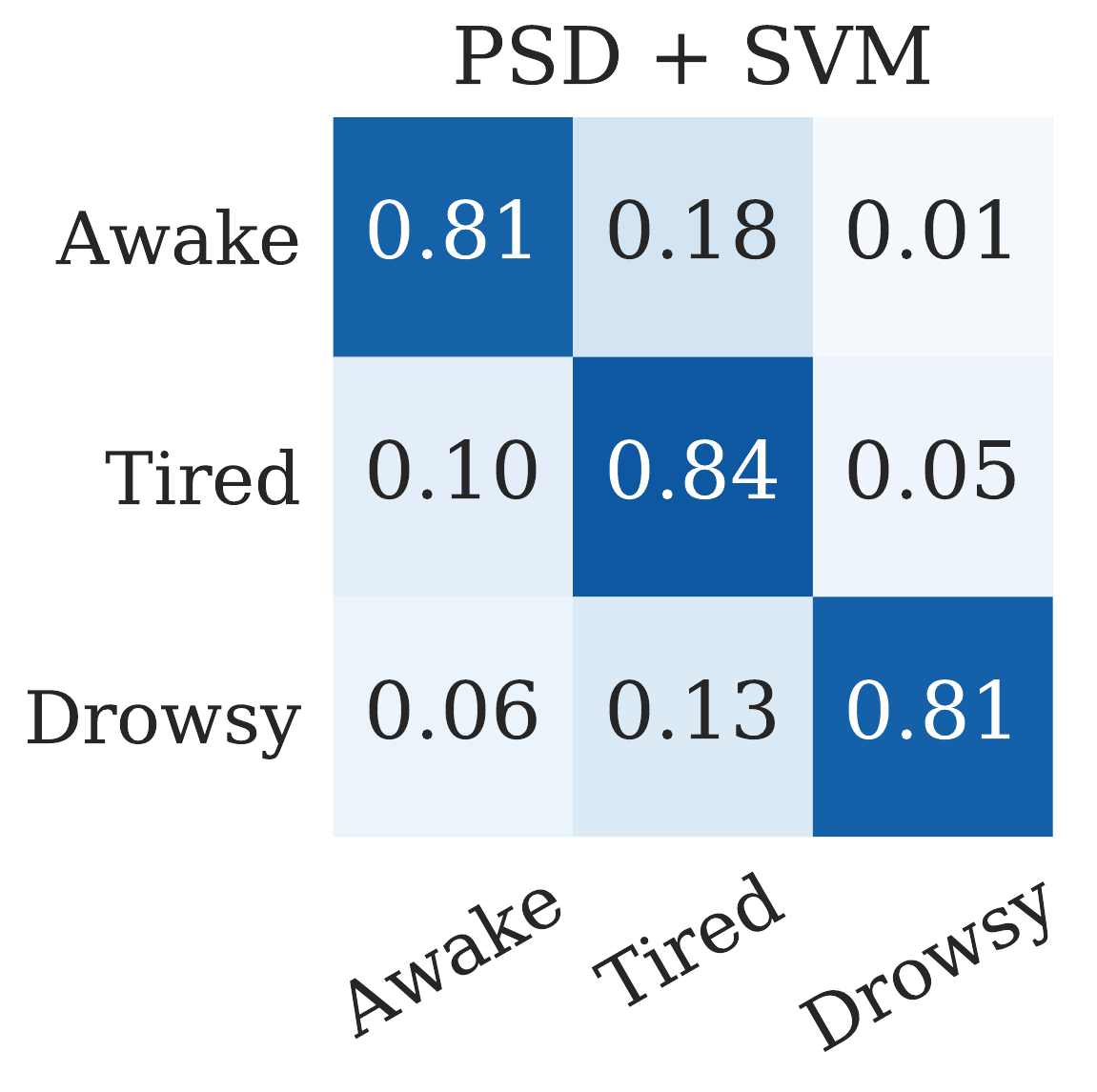}
		\includegraphics[width=.12\linewidth]{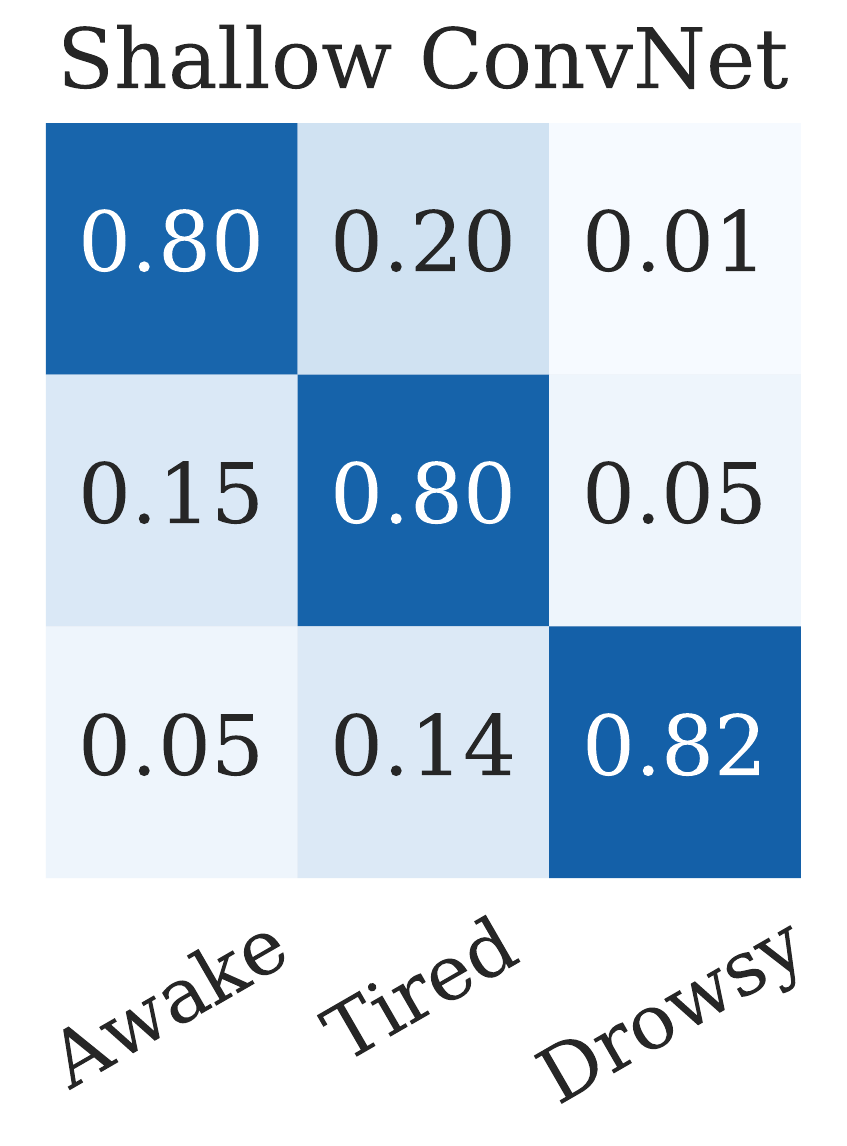}
		\includegraphics[width=.12\linewidth]{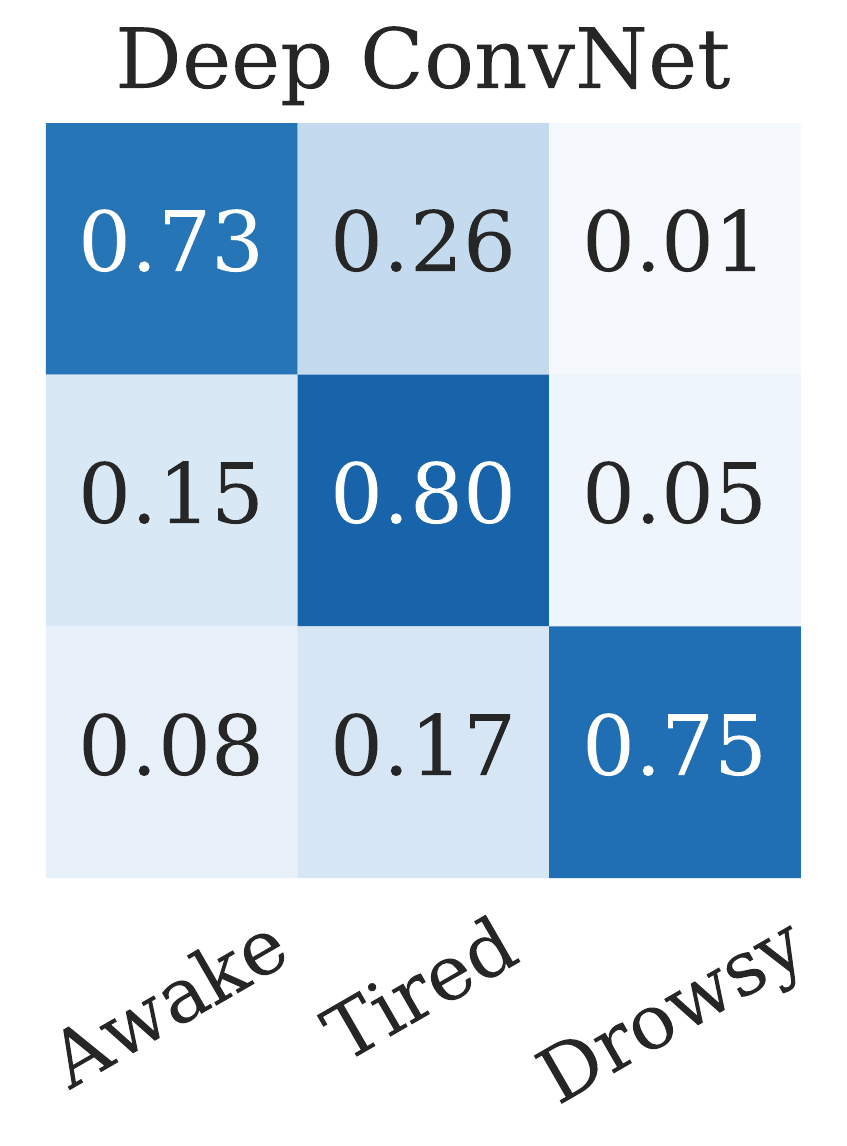}
		\includegraphics[width=.12\linewidth]{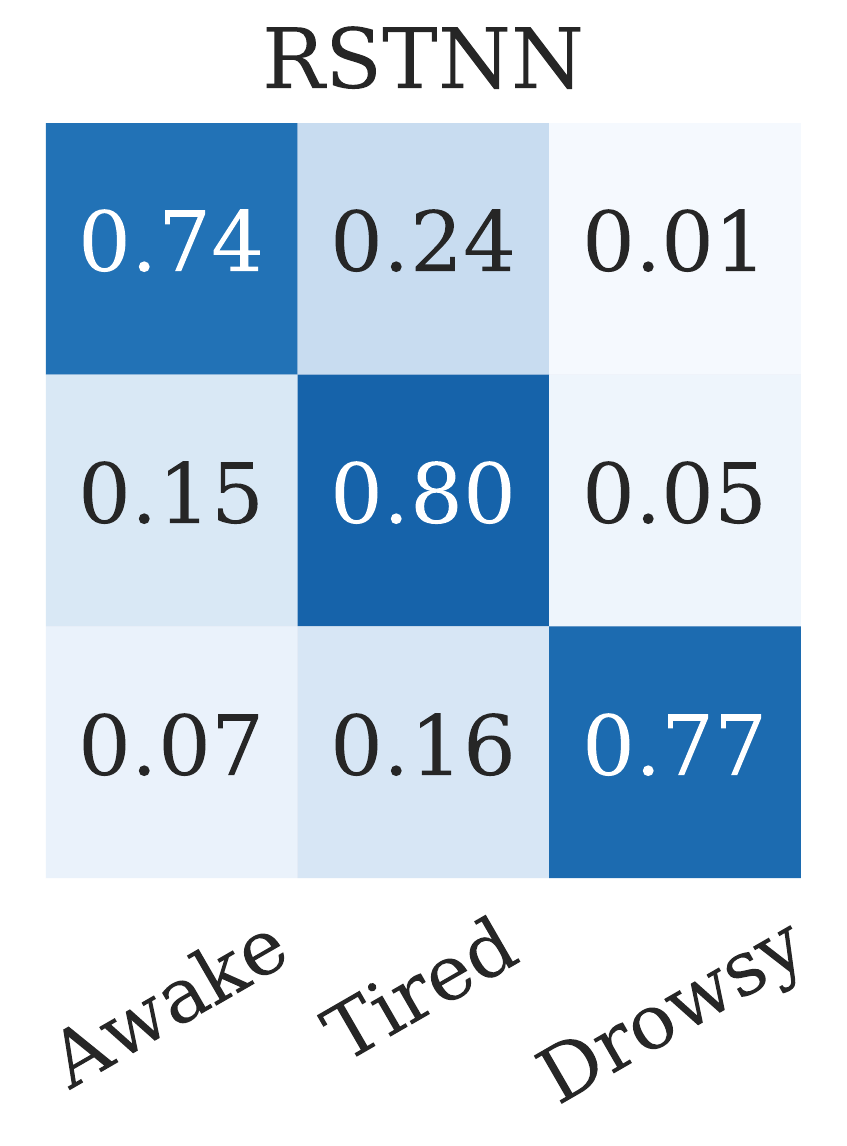}
		\includegraphics[width=.12\linewidth]{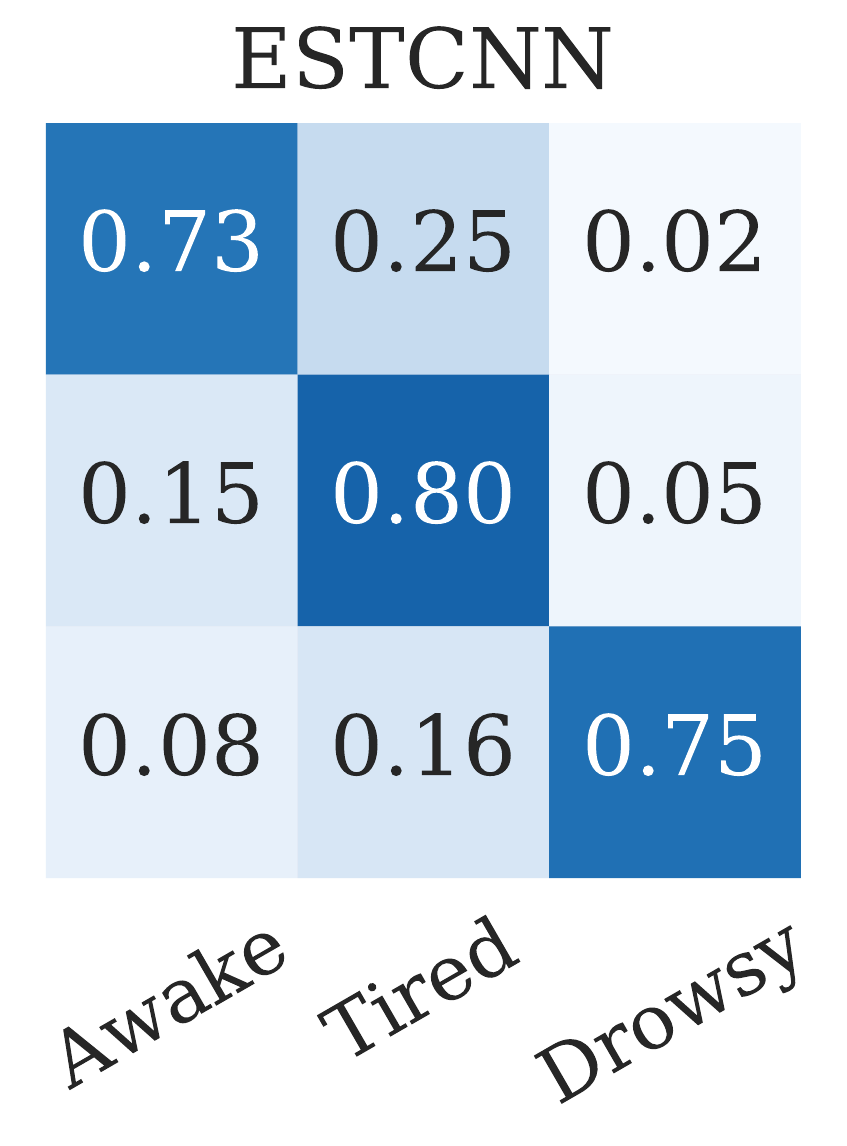}
		\includegraphics[width=.12\linewidth]{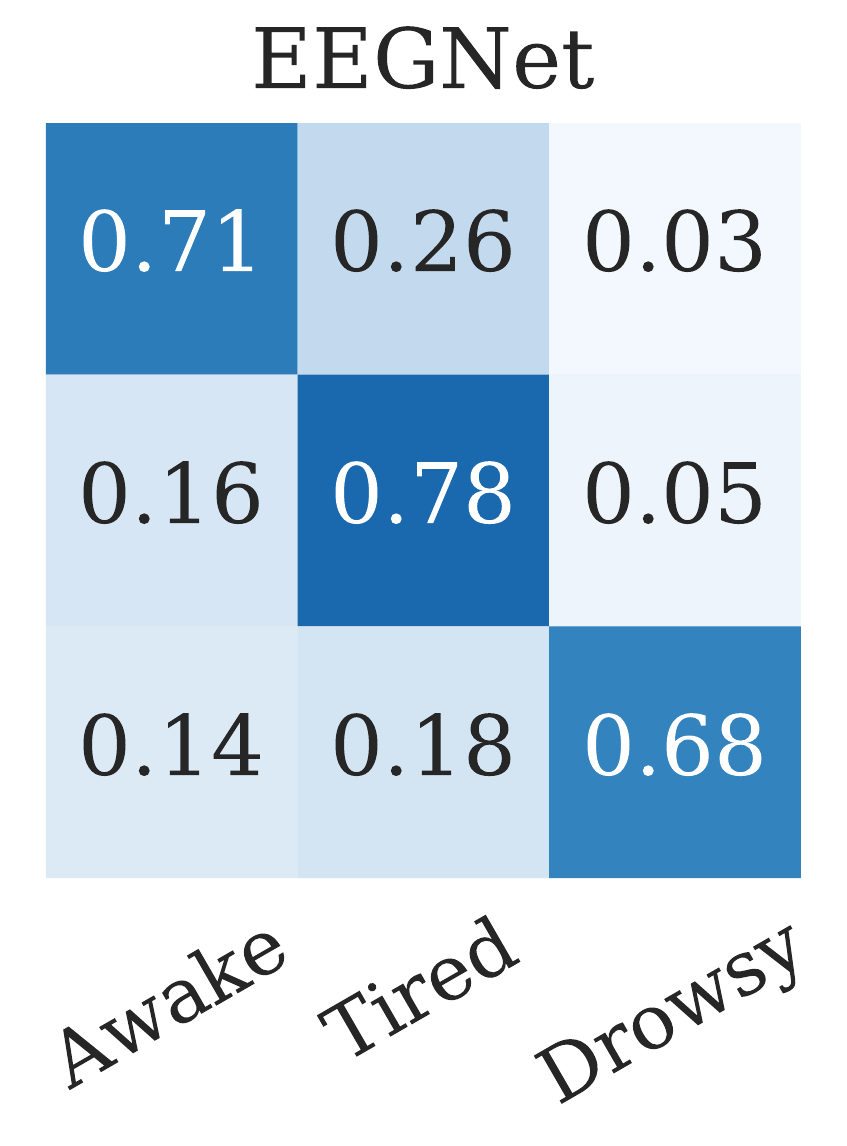}
		\includegraphics[width=.165\linewidth]{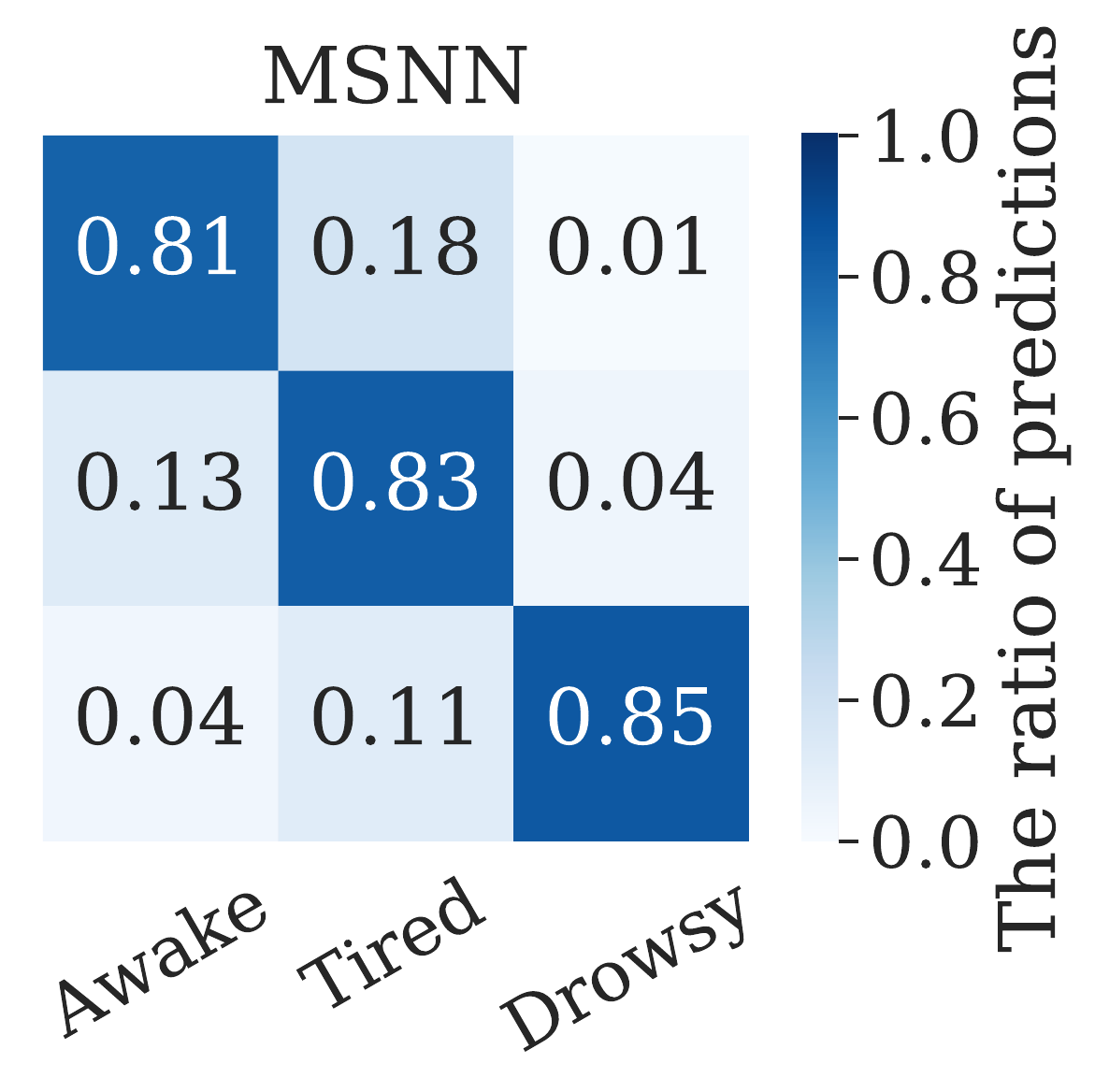}
		\caption{Normalized averaged confusion matrices estimated by comparable baselines and the proposed method using the SEED-VIG dataset \cite{zheng2017multimodal}.}
		\label{subfig: normalized_confusion_matrix}
	\end{subfigure}\vspace{5pt}
	\begin{subfigure}{1\linewidth}
	\centering
	\includegraphics[width=.85\linewidth]{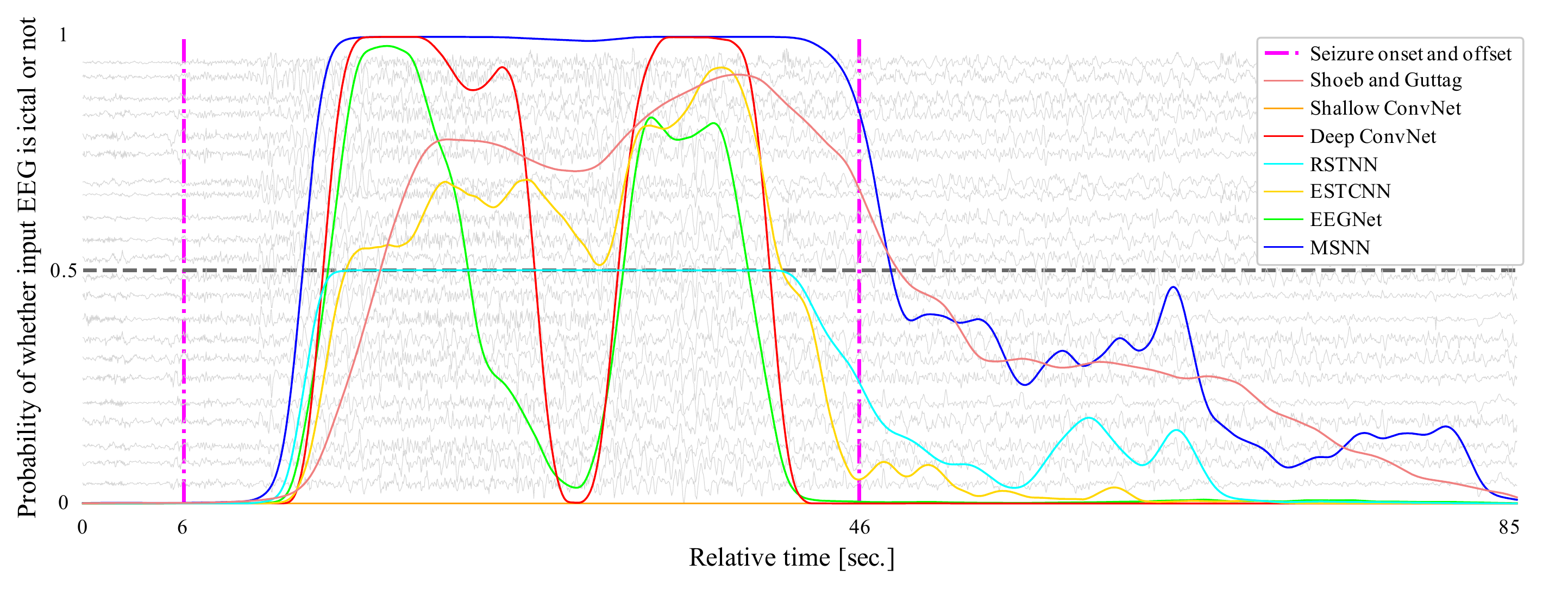}\vskip-.05in
	\caption{Changes of probabilities estimated by comparable baselines, and the proposed method. These plots demonstrate the probability of whether input EEG is ictal or not. Two dot-dashed lines (magenta) denote the seizure onset and ending, respectively, labeled by Shoeb \cite{shoeb2009application}.}
	\label{subfig: seizure_probability}
\end{subfigure}
	\caption{Investigation of learned weights (Fig. \ref{subfig: activation_patterns}) and represented features (Fig. \ref{subfig: tsne_plotts}), and inspection of the practical usage of the proposed network (Fig. \ref{subfig: seizure_probability} and \ref{subfig: normalized_confusion_matrix}).}
	\label{fig: investigation}
\end{figure*}

\subsection{Activation Patterns}
\label{subsec: activation patterns}
Earlier, Haufe \etal~\cite{haufe2014interpretation} proposed an \emph{activation pattern} which is based on a \emph{forward-backward modeling} in signal processing. The activation pattern method \cite{haufe2014interpretation} provides a way to interpret weight matrices in multivariate neuroimaging, as presented in the signal processing literature.

The proposed method, clearly, decodes the input EEG signal to the corresponding label, \ie, inferring a user's intention or condition from an observed EEG pattern. Therefore, it is a backward process computational model. Hence, for a concrete and meaningful understanding of learned layers, it is essential to reverse this backward process model to a forward process. Finally, in this work, we estimated and visualized the activation patterns of the learned weights shown in Fig. \ref{subfig: activation_patterns}. We extracted the spatial convolutions of Shallow ConvNet \cite{schirrmeister2017deep}, Deep ConvNet \cite{schirrmeister2017deep}, RSTNN \cite{ko2018deep}, EEGNet \cite{lawhern2018eegnet}, and the proposed model. Then, we estimated activation patterns and visualized them in a topological manner. We do not estimate ESTCNN \cite{gao2019eeg} activation patterns because the ESTCNN \cite{gao2019eeg} does not have any spatial feature representation layers and those visualized patterns are estimated by the first subject's first fold data in the GIST-MI dataset \cite{cho2017eeg}. Finally, we normalized the activation patterns in [0, 1] range before visualization.

In this investigation, we observed right-lateralized brain activation/deactivation patterns, and the same patterns in the left hemisphere when a user imagined the movement of left-hand and right-hand respectively. Furthermore, the proposed model shows relatively clearer patterns than the other models, thus, we can conclude that our method thoroughly represents input EEG signal spatial features.
\subsection{Discriminative Power of EEG Representations}
\label{subsec: eeg representations}
To validate the representation ability of the proposed network, we plotted t-SNE transformed learned features shown in Fig. \ref{subfig: tsne_plotts}. Specifically, we exhibited extracted features from test SSVEP EEG samples from the first, second, and third spatio-spectral-temporal feature representation layers, \ie, $\mathbf{f}_1^\text{SST}$, $\mathbf{f}_2^\text{SST}$, and $\mathbf{f}_3^\text{SST}$ (first three figures in Fig. \ref{subfig: tsne_plotts}). Then, we also depicted the final learned feature, \ie, $\mathcal{G}(\mathbf{f}_\text{concat}^\text{SST})$. These intermediate features $\mathbf{f}_1^\text{SST}$, $\mathbf{f}_2^\text{SST}$, and $\mathbf{f}_3^\text{SST}$ are temporally pooled just for visualization like $\mathcal{G}(\mathbf{f}_\text{concat}^\text{SST})$. We used the first subject's first session data in the KU-SSVEP dataset \cite{lee2019eeg}, and used a learning rate of 200, a perplexity of 10 for the t-SNE calculation, and visualization. 

From these visualized represented features, we could observe that $\mathcal{G}(\mathbf{f}_\text{concat}^\text{SST})$ is more class-discriminative than the other intermediate features. Additionally, we observed a trend, which demonstrated that a feature learned by a deeper layer is more disentangled than others learned by shallower layers.

\subsection{Mental Fatigue Classification}
\label{subsec: mental fatigue classification}
For the application analysis of drowsiness detection, we visualized confusion matrices that were estimated by the experimental results of the SEED-VIG dataset \cite{zheng2017multimodal} in Fig. \ref{subfig: normalized_confusion_matrix}. Because the labels that identify the mental status were decided using the PERCLOS levels \cite{zheng2017multimodal}, the label at the boundary of the two classes may not be accurate. In this respect, we can conclude that the proposed method is useful for drowsiness state detection because false detections predicted by the proposed method are mostly at the boundaries between classes, \eg, the `awake' \emph{vs.} `tired' or `tired' \emph{vs.} `drowsy' case. In addition, for practical application, it is essential to detect the drowsy state accurately to avoid unexpected situations, such as a car accident. The proposed method achieved the highest and most promising result for detecting drowsiness among other baselines, \ie, it achieved the highest precision score for identifying the drowsy state. Therefore, we can also expect that our proposed method can be applied in real-world situations.

\subsection{Early Seizure Detection}
\label{subsec: early seizure detection}
Early detection \cite{lee2017early} of seizures is one of the most important potential practical applications for this work. Hence, we also validated tthe benefits of the proposed method in early seizure detection. Specifically, in the training phase, the MSNN was trained using normal and ictal EEG samples with binary labels (\eg, 0: normal and 1: seizure) similar to a conventional training framework. In the testing phase, we input the EEG samples using a sliding window with a 1/256 stride. Then, we observed the change in the output probability values to determine the character of the input (normal or ictal).

Additionally, we visualized these changes in Fig. \ref{subfig: seizure_probability} (We used the first subject's third EEG trial in the CHB-MIT dataset \cite{shoeb2009application} for the visualization). In Fig. \ref{subfig: seizure_probability}, magenta-colored dot-dashed lines denote the seizure onset and offset. Colored solid lines denote the probability change of various methods. In this visualization, we observed that the proposed method is more stable for detecting seizures. Specifically, the proposed method detects the seizure EEG signal as a seizure state with a strong probability (almost 1), whereas the other methods have low confidence values (Shoeb and Guttag \cite{shoeb2010application}'s method and ESTCNN \cite{gao2019eeg}) or even make incorrect decisions regarding the seizure state (Shallow ConvNet \cite{schirrmeister2017deep}, Deep ConvNet \cite{schirrmeister2017deep}, RSTNN \cite{ko2018deep}, and EEGNet \cite{lawhern2018eegnet}).

\section{Conclusion}
\label{sec: conclusion}
In this work, we proposed a novel and compact deep multi-scale neural network which can learn multi-scale EEG signal features. In our experiments, we validated our novel architecture's effectiveness over diverse EEG paradigms, MI, SSVEP, seizure, and drowsy EEG signals. Furthermore, we inspected the relevance scores to demonstrate the benefits of the multi-scale feature extraction ability, investigated activation pattern maps to understand what types of neurophysiological phenomena were learned by our CNN model, and visualized the t-SNE of learned features to examine the ability of our method to differentiate feature classes. Finally, we also demonstrated that the proposed method can be used for precise drowsiness detection and early seizure detection. In all these respects, we concluded that the proposed deep multi-scale neural network offers significant potential for interpreting EEG signals. Additionally, because the proposed network is clearly \emph{generalizable} to various EEG paradigms, it is expected to have promising benefits that can apply to neural architecture search methods \cite{rapaport2019eegnas}, thereby making a deep learning-based BCI adaptable to different paradigms.

From a practical standpoint, many limitations remain with regard to the inter-subject variation \cite{jayaram2016transfer} in performance. In the present work, we experimented in a subject-dependent manner. In general use, it is important for a BCI system to be useful for any subject operating in a subject-independent way. Thus, in the future, we will focus on developing a subject-neutral multi-paradigm BCI system using adversarial learning \cite{ganin2016domain, jeon2019domain} or other learning strategies \cite{li2017learning}.

\appendices

\section*{Acknowledgment}
This work was supported by Institute for Information \& Communications Technology Promotion (IITP) grant funded by the Korea government (No. 2017-0-00451, Development of BCI based Brain and Cognitive Computing Technology for Recognizing User's Intentions using Deep Learning).

\ifCLASSOPTIONcaptionsoff
  \newpage
\fi

\bibliographystyle{IEEEtran}
\bibliography{main}


%
%
%

\end{document}